# Secure Transmission with Different Security Requirements Based on Covert Communication and Information-Theoretic Security in Presence of Friendly Jammer

1. Pooya Baee, Faculty of New Technologies Engineering, Amol University of Special Modern Technologies,Iran,pooya_baee@outlook.com
2. Farid Samsami khodadad, Faculty of New Technologies Engineering, Amol University of Special Modern Technologies,Iran, samsami@ausmt.ac.ir
3. Moslem Forouzesh, Faculty of Electrical and Computer Engineering, Trabiat Modares University,Iran, M.forouzesh@modares.ac.ir

*Abstract:* **In this paper, we investigate joint information-theoretic security and covert communication on a network in the presence of a single transmitter (Alice), a friendly jammer, a single untrusted user, two legitimate users, and a single warden of the channel (Willie). In the considered network, one of the authorized users, Bob, needs a secure and covert communication, and therefore his message must be sent securely, and at the same time, the existence of his communication with the transmitter should not be detected by the channel's warden, Willie, Meanwhile, another authorized user, Carol, needs covert communication. The purpose of secure communication is to prevent the message being decoded by the untrusted user who is present on the network, which leads us to use one of the physical layer security methods, named the secure transmission of information theory. In some cases, in addition to protecting the content of the message, it is important for the user that the existence of the transmission not being detected by an adversary, which leads us to covert communication. In the proposed network model, it is assumed that for covert communication requirements, Alice will not send any messages to legitimate users in one time slot and in another time slot will send to them both (Bob and Carol). One of the main challenges in covert communication is low transmission rate, because we have to reduce the transmission power such that the main message get hide in background noise. In the proposed network, a jammer is used to destroy the eavesdropper and warden's channel. In order to eliminate the interference created by the jammer, a friendly jammer has been used in the network. In this study, in addition to using joint security of information theory and covert communication, we examine the average transmission rate according to the limitations and requirements of covert communication and the requested quality of service by users, by creating artificial noise in the network which is generated by a friendly jammer.**

## 1. Introduction

Given the increasing use of wireless networks in both military and civilian applications, creating security in these types of networks, which due to their broadcast nature are exposed to various attacks such as eavesdropping and traffic analysis, is an important challenge. Therefore, sharing confidential information reliably in the presence of enemies is very important. Enemies may carry out various attacks to gain unauthorized access to or change the information or even to disrupt network's performance [1].

Security methods such as encryption methods used in the higher layers of the network are not completely confidential due to the increasing development of computing devices, and it is possible that the eavesdropper will access the content of our message by breaking the password (even after several years) and this will not be favorable in some situations with a high security approach. Due to the symmetric encryption method, such as the data encryption standard, a common dedicated key is usually shared between two users. If these two users do not have the dedicated key, a separate protected channel is required to exchange the shared key. Instead of using an additional channel, physical layer methods are used to share the secret key. The use of physical layer protection schemes makes it difficult for enemies to decrypt transmitted information [2].

In older security methods, it was assumed that the computing power of the eavesdropper was low and therefore eavesdropper is unable to detect the secret key. But as we know, equipment is growing extremely in terms of computational power. Thus, information theory security methods are promising ideas for wireless telecommunications security in which additional security methods are not used [3].

Traditional security methods offered to protect against eavesdropping by encrypting, ensuring the integrity of the message in the air. However, in recent years it has been shown that even enhanced encryption methods can be defeated by eavesdroppers [4].

Physical layer security methods using the dynamic specifications of wireless media minimize the information obtained by the channel's eavesdropper, while these methods primarily do not provide covert communication between two users [5].

In general, secure transmission methods in the physical layer are divided into 5 main categories: secure information-theoretic capacity, channel methods, coding, power methods, and signal detection methods [2].

Secure transmission at the physical layer is usually modeled using the concept of eavesdropping channel. In this model, the transmitter tries to have secure communication with the intended recipient, so that the eavesdropper cannot receive confidential messages. According to Wyner, to provide security without the need for cryptography, the signal received by the eavesdropper must be a weaker, less detectable sample than the signal received by the main receiver, i.e. the eavesdropper's channel must be noisier than the main channel [6].

On the other hand, there are situations where transmission between the transmitter and receiver needs to be done covertly. In other words, the goal is to hide the existence of communication between the transmitter and receiver from an adversary. Applications for covert communications include: in military communications, it is sometimes necessary to hide the transmitter's activity in sending data in a geographical area from the eavesdroppers or enemies, because the enemy may take action if he understands the sending activity [7].

In addition to protecting the content of the message, covert communications, commonly referred to as low-detection communication, attempt to provide a wireless transmission between two users, which also ensures that the transmission is less likely to be detected by an eavesdropper. Such communications are ideal for the politicians and military applications who are interested in keeping their communications over wireless media anonymous. Covert communication has attracted a lot of attention in recent years and has emerged as a new method in the form of wireless communication security [8] , [9].

In [10] joint information theoretic security and covert communication was investigated at which users with different security needs are presented at network. In [11] and [12] it has been proven that for secure transmission without encryption, a positive transmission rate will be available if the adversary is unsure of the received noise power at its receiver. It is also shown in [13] that if we get help from a jammer, a positive transmission rate will be achievable for us.

The full-duplex receiver strategy has been used in many studies, including [13] and [14], in which the receiver can send jamming signals at the same time as receiving a message from the source to mislead the adversary in same frequency band.

In [5], [15] and [16] the probabilities and conditions of covert communication in a static Gaussian fading channel using artificial noise generation (AN) which is generated by a full-duplex receiver have been investigated, in [5] at which the desired level of covertness can be achieved by controlling the random power of artificial noise.

In [17] improving security by cooperative jamming which is achieved by disturbance signals sent by users or auxiliary relay nodes has been investigated. In [18], the physical layer security in the presence of an adversary who can change his state from eavesdropping to disruptive mode has been investigated. In the first case, called eavesdropping, the enemy tries to eavesdrop on the authorized user's channel, and in the second case, called jamming; the enemy sends a distortion signal (artificial noise) to mislead the main receiver.

In [19] it has been shown that jamming can significantly increase the rate of covert communication, and if we reduce the interference at the legitimate receiver (Bob), the transmission rate in the covert communication is even more than this. In [19], to reduce the interference in the legal receiver, a multi-antenna jammer that uses null space beamforming, and a multi-antenna transmitter with three-dimensional antennas that can radiate to the desired receiver and potentially far from the adversary is being employed.

In general, jamming against eavesdroppers was introduced because the form of radiation was not responsible for the secure transmission to destroy the eavesdropper channel. Artificial noise generation is a method of disrupting eavesdropper in a network that can sometimes be done by the transmitter [20] or even the legal receiver [21], but these disruption methods reduce network's efficiency due to channel conditions and severe self-interference [22]. Instead, a more effective way is to use a friendly jammer in the network. This friendly jammer emits artificial noise to reduce the signal-to-interference plus noise ratio (SINR) of an eavesdropper, at the expense of additional power and interface costs [23].

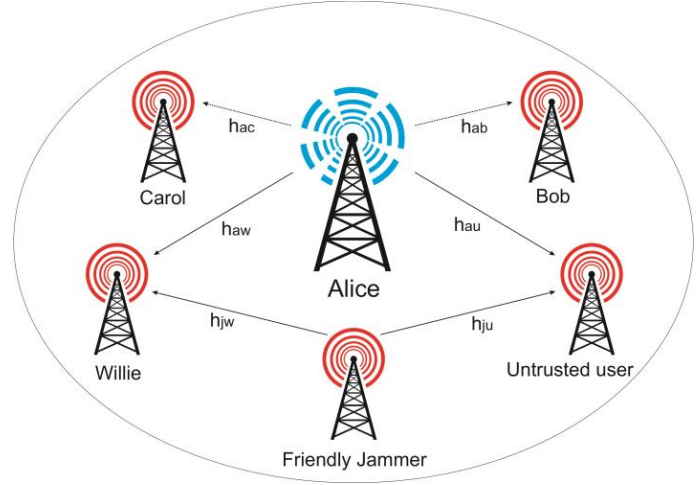

**Figure 1.** Proposed system model: Secure and covert transmission in the presence of a friendly jammer.

We briefly describe our main goals in this article as follows:

- In this model, we will study a system in which two types of authorized users with different levels of security are present. One of these authorized users needs secure and covert communication and the other only needs covert communication.

- In this study, in order to increase the average rate, we use an external friendly jammer and investigate the effect of its presence.

- In this paper, we will investigate the influence of imperfect information about Willie's location on network's performance.

The configuration of this article is described below. In the second part we'll introduce system model and describe communication scenario and also we will examine the secure and covert communications requirements. In the third section, we will discuss the optimization problem and in fourth section we will find a solution for it. In fifth section, the scenario of information uncertainty from the eavesdropper's location will be examined. In section six, we present numerical results and in section seven of this article, we will conclude.

## 2. System model

### I. Transmission scenario and assumptions

The proposed system model is shown in Figure 1, which includes a transmitter (Alice), an untrusted user, two authorized users (Bob) and (Carol), a warden (Willie), and a friendly jammer.

It should be noted that the nature of the untrusted user is not known to the network and therefore he can be an eavesdropper. The distances between Alice and Bob, Alice and Carol, Alice and Willie, Alice and untrusted user, friendly jammer and Willie, and friendly jammer and untrusted users are $d_{ab}, d_{ac}, d_{aw}, d_{au}, d_{jw}, d_{ju}$ respectively. The channel fading coefficients between Alice to Bob, Alice to Carol, Alice to Willie, Alice to untrusted user, friendly jammer to Willie, and friendly jammer to untrusted user are $h_{ab}, h_{ac}, h_{aw}, h_{au}, h_{jw}, h_{ju}$ respectively and these channels have circularly symmetric complex Gaussian distribution with zero mean and unit variance. In this paper, it is assumed that all channel coefficients are constant in one time slot and changes in another period and are independent of each other.

In the proposed network model, it is assumed that for covert communication requirements, Alice will not send any messages to two users who need secure message and covert communication at one time slot, and will send to both of them at another time slot. Since the eavesdropper is passive, it is assumed that the eavesdropper's channel state information (CSI) is not available on the network. The reason why Willie is passive is that if he is active, he will help to covert communication, which is not desirable for him. Also in this system, a discrete time channel with $Q$ time interval is considered, which the length of each of these intervals is $n$ symbol. Hence the transmit signal from Alice to Carol and Bob, friendly jammer to untrusted user and Willie in a time slot are $\mathbf{x}_c = [x_c^1, x_c^2, ..., x_c^n]$, $\mathbf{x}_b = [x_b^1, x_b^2, ..., x_b^n]$, $\mathbf{x}_{ju} = [x_{ju}^1, x_{ju}^2, ..., x_{ju}^n]$ and $\mathbf{x}_{jw} = [x_{jw}^1, x_{jw}^2, ..., x_{jw}^n]$, respectively, where $n$ is the total number of symbols in a time slot. It should be noted that Alice does not send $\mathbf{x}_b$ and $\mathbf{x}_c$ continuously due to covert communication requirements.

### II. Information Theoretic Security

As mentioned, Alice's transmitter antenna will not send any messages to either of the two authorized users within a certain time slot, and at another time slot he will send to both authorized users of the network, Bob, who needs secure and covert communication, and Carol, who needs covert communication. It should also be noted that, as mentioned earlier, a friendly jammer has been used in the network in order to destroy the network's eavesdropper (Willie) channel, and therefore the noise generated by the friendly jammer is known to legitimate network users, Bob and Carol, and can ignore it in their receiver and according to the assumptions, the signal vector received in each node $m$ of this network (Bob, Carol, untrusted user and Willie) in each time slot is as follows:

$$y_m = \begin{cases} \dfrac{\sqrt{p_j} h_{jm} x_j}{d_{jm}^{\alpha/2}} + \mathbf{N_m}, & \psi_0 \\ \dfrac{\sqrt{p_{ab}} h_{am} x_b}{d_{am}^{\alpha/2}} + \dfrac{\sqrt{p_{ac}} h_{am} x_c}{d_{am}^{\alpha/2}} + \dfrac{\sqrt{p_j} h_{jm} x_j}{d_{jm}^{\alpha/2}} + \mathbf{N_m}, & \psi_1 \end{cases} \quad (1)$$

where $p_j$, $p_{ab}$ and $p_{ac}$ are friendly jammer power, Alice's transmit power for Bob and Carol, respectively. $\alpha$ is the pass-loss exponent, and $\mathbf{N}_m \sim CN(0, \sigma_m^2)$ represents the received noise power at $m$. In the above equation, the symbol

$\psi_0$ indicates that Alice doesn't have any transmission to authorized receivers in the network who need secure and covert communication, while $\psi_1$ indicates the transmission of message within a specified time slot. It should be noted that since the Alice's transmitter is a single antenna, the power transmitted to each of the authorized receivers of the network is a percentage of the total power of the Alice antenna, and to simplify the relationships, the power transmitted to Carol can be considered as $p_{ac} = 1 - p_{ab}$ and it can also be said that the total transmission power is limited in $P_{max}$, which is a common assumption in [24]. Hence we consider the friendly jammer power and antenna power of Alice's transmitter for each time slot as
$\begin{cases} p_j P_{jmax} & \psi_1 \\ (p_{ab}+p_{ac})P_{max} = P_{max} & \psi_1 \end{cases}$ and $\begin{cases} p_j P_{jmax} & \psi_0 \\ 0 & \psi_0 \end{cases}$
respectively, where $p_{ab} \in [0,1]$ and $p_j \in [0,1]$, which this assumption has been used in relationships. Next, we consider the variable $\gamma_m = \frac{P_{max}|h_{am}|^2}{\sigma_m^2 d_{am}^\alpha}$ to calculate the SINR, and to simplify the relationships. Therefore, the SINR at the Bob's receiver is as follows:

$$\gamma_B^\ell = \begin{cases} 0 & \psi_0 \\ \frac{p_{ab}\gamma_b}{p_{ac}\gamma_b + 1} & \psi_1 \end{cases} \quad (2)$$

and also the SINR at the untrusted user's receiver is as follows:

$$\gamma_U^\ell = \begin{cases} p_j \gamma_u & \psi_0 \\ \frac{p_{ab}\gamma_u}{p_{ac}\gamma_u + p_j\gamma_j + 1} & \psi_1 \end{cases} \quad (3)$$

where $\gamma_j = \frac{P_{jmax}|h_{ju}|^2}{\sigma_u^2 d_{ju}^\alpha}$, and Finally, the SINR for the user who needs covert communication, i.e. Carol, is as follows:

$$\gamma_C^\ell = \begin{cases} 0 & \psi_0 \\ \frac{p_{ac}\gamma_c}{p_{ab}\gamma_c + 1} & \psi_1 \end{cases} \quad (4)$$

Therefore, the secure rate of information theory on the Bob receiver will be calculated as follows:

$$R_{sec}(\mathbf{P}) = [\log_2(1+\gamma_B^\ell) - \log_2(1+\gamma_U^\ell)]^+ \quad (5)$$

where $[x]^+ = \max\{x,0\}$, and $\mathbf{P}$ matrix contains $p_{ab}$ and $p_j$ components.

### III. Covert Communication requirements

The eavesdropper (Willie) criterion for detecting the existence of communication between two nodes is based on the analysis of the signal strength received at its receiver, and based on this analyze he will decide whether or not a message have been sent from the transmitter (Alice) to the his intended recipient (Carol or Bob). In short, Willie is faced with a binary hypothesis. When Alice has sent a message to the intended receiver and Willie decides based on the received signal strength on the absence of transmission, he will have a missed detection with a probability of $p_{MD}$. Also, when the Alice has not sent a message to the intended recipient but Willie decides on the existence of communication between two nodes, he will have a false alarm with a probability of $p_{FA}$. It should be noted that the necessary condition for establishing a covert communication between two nodes is that the following condition is met [7]:

$$\text{for any } \varepsilon \geq 0, p_{MD} + p_{FA} \geq 1 - \varepsilon \text{ as } n \to \infty \quad (6)$$

Also, the optimal decision-making rule at Willie to reduce the detection error will be as follows [7]:

$$\begin{cases} \dfrac{Y_w}{n} < \theta & \Psi_0 \\ \dfrac{Y_w}{n} > \theta & \Psi_1 \end{cases} \quad (7)$$

where $Y_w = \sum_{\ell=1}^{n} |y_w^\ell|^2$, is the total received power at Willie in each time slot and $\theta$ is decision threshold at Willie. In the following, we will calculate the probabilities of false alarm (FA) and missed detection (MD).

## IV. False alarm and Missed detection probabilities

The probability of missed detection and false alarm can be calculated as follows:

$$p_{FA} = P\left(\dfrac{Y_w}{n} > \theta \Big| \Psi_0\right) \quad (8)$$

$$p_{MD} = P\left(\dfrac{Y_w}{n} < \theta \Big| \Psi_1\right) \quad (9)$$

In order to calculate the above probabilities, we will need the probability distribution function of the random variable $\gamma_w^\ell$. It is assumed that the fading in this network has a Rayleigh distribution and therefore each signal symbol received at the eavesdropper's receiver (Willie) i.e., $y_w^\ell$ has circular complex Gaussian statistical distribution as follows:

$$y_w^\ell \sim CN(0, \sigma_w^2 + \gamma_w^\ell) \quad (10)$$

which

$$\gamma_w^\ell = \begin{cases} p_j P_{j\max} d_{jw}^{-\alpha} |h_{jw}|^2 & \Psi_0 \\ P_{\max} d_{aw}^{-\alpha} |h_{aw}|^2 + p_j P_{j\max} d_{jw}^{-\alpha} |h_{jw}|^2 & \Psi_1 \end{cases} \quad (11)$$

and the probability Density Function (PDF) of $\gamma_w^\ell$ is:

$$f_\Psi(\gamma_w^\ell) = \begin{cases} \dfrac{1}{\lambda_1} \times \left(e^{-\dfrac{\gamma_w^\ell}{\lambda_1}}\right) & \gamma_w^\ell > 0, \Psi_0 \\ \dfrac{1}{\lambda_2 - \lambda_1} \times \left(e^{-\dfrac{\gamma_w^\ell}{\lambda_2}} - e^{-\dfrac{\gamma_w^\ell}{\lambda_1}}\right) & \gamma_w^\ell > 0, \Psi_1 \end{cases} \quad (12)$$

where $\lambda_1 = p_j P_{j\max} d_{jw}^{-\alpha}$ and $\lambda_2 = P_{\max} d_{aw}^{-\alpha}$.

As we know, the sum of $n$ random variables with a chi-square distribution with two degrees of freedom will have a chi-square distribution with $2n$ degrees of freedom. According to the above, $Y_w$ have chi-square distribution with $2n$ degrees of freedom and so

$$p_{FA} = P\left(\dfrac{Y_w}{n} > \theta \Big| \psi_0\right) = P\left((\sigma_w^2 + \gamma_w^\ell)\dfrac{\chi_{2n}^2}{n} > \theta \Big| \psi_0\right) \quad (13)$$

$$p_{MD} = P\left(\dfrac{Y_w}{n} < \theta \Big| \psi_1\right) = P\left((\sigma_w^2 + \gamma_w^\ell)\dfrac{\chi_{2n}^2}{n} < \theta \Big| \psi_1\right) \quad (14)$$

which $\chi_{2n}^2$ is a random variable of chi-square with 2n degree of freedom. If we have $n \to \infty$, and we consider the probability that the condition of the channel is such that the covert communication is completed, according to the law of large numbers, $\chi_{2n}^2$ converges to 1, and according to the Lebesgue's Dominated Convergence Theorem, when we have $n \to \infty$, we can replace $\dfrac{\chi_{2n}^2}{n}$ with 1. Using above results, we have the following probabilities of false alarm and missed detection as follows:

$$p_{FA} = \begin{cases} \left(e^{-\dfrac{(\theta - \sigma_w^2)}{\lambda_1}}\right) & \theta - \sigma_w^2 \geq 0 \\ 1 & \theta - \sigma_w^2 < 0 \end{cases} \quad (15)$$

$$p_{MD} = \begin{cases} \frac{1}{(\lambda_2 - \lambda_1)} \times \left( \lambda_2 \times \left[ 1 - e^{-\frac{(\theta - \sigma_w^2)}{\lambda_2}} \right] - \ldots \right. \\ \left. \lambda_1 \times \left[ 1 - e^{-\frac{(\theta - \sigma_w^2)}{\lambda_1}} \right] \right) & \theta - \sigma_w^2 \geq 0 \\ 0 & \theta - \sigma_w^2 < 0 \end{cases} \quad (16)$$

### V. *Optimal decision threshold for Willie*

Since the object of the adversary (Willie) is to minimize $P_{FA} + P_{MD}$, the adversary will never choose the value of $\theta$ as $\theta < \sigma_w^2$, because in that case Willie will be face to $P_{FA} + P_{MD} = 1$. So we will choose an expression for the case of $\theta > \sigma_w^2$. In view of the above, in order to obtain the optimal threshold for the decision of Willie i.e. $\theta_{op}$, we consider the expression $\frac{\partial (P_{FA} + P_{MD})}{\partial \theta} = 0$ and the $\theta_{op}$ will be obtained as follows:

$$\theta_{op} = \frac{(\lambda_2 \times \lambda_1)}{(\lambda_2 - \lambda_1)} \times Ln\left(\frac{\lambda_2}{\lambda_1}\right) + \sigma_w^2 \quad (17)$$

### 3. Optimization problem

In this section, in order to estimate the proposed network model, we propose the optimization problem, in which the main goal is to maximize the average rate with respect to power constraints, quality of service requested by users, and covert communication requirements. In the time slots in which Alice has transmission to Bob and Carol, the total rate is obtained as follows:

$$R_{\sec T}(p_{ab}, p_j) = \log_2\left(1 + \frac{p_{ac}\gamma_c}{p_{ab}\gamma_c + 1}\right) + \ldots$$
$$\left[\log_2\left(1 + \frac{p_{ab}\gamma_b}{p_{ac}\gamma_b + 1}\right) - \ldots \right. \\ \left. \log_2\left(1 + \frac{p_{ab}\gamma_u}{p_{ac}\gamma_u + p_j\gamma_j + 1}\right)\right] \quad (18)$$

as a result, we will define the optimization problem as follows

$$\max_{p_{ab}, p_j, t} \log_2\left(1 + \frac{(1 - p_{ab})\gamma_c}{p_{ab}\gamma_c + 1}\right) + \ldots$$
$$\left[\log_2\left(1 + \frac{p_{ab}\gamma_b}{(1 - p_{ab})\gamma_b + 1}\right) - \ldots \right. \\ \left. \log_2\left(1 + \frac{p_{ab}\gamma_u}{(1 - p_{ab})\gamma_u + p_j\gamma_j + 1}\right)\right] \quad (19)$$

s.t.
$$0 \leq p_j \leq 1 \quad (19.a)$$
$$0 \leq p_{ab} \leq 1 \quad (19.b)$$
$$\left[\log_2(1 + \gamma_B^\ell) - \log_2(1 + \gamma_U^\ell)\right] \geq R_{Bob}^{\min} \quad (19.c)$$
$$\log_2\left(1 + \frac{p_{ac}\gamma_c}{p_{ab}\gamma_c + 1}\right) \geq R_{Carol}^{\min} \quad (19.d)$$
$$\min(P_{FA} + P_{MD}) \geq 1 - \varepsilon \quad (19.e)$$

by replacing $\theta_{op}$ in constraint (19.e) we will have this as follows:

$$\left(\frac{\lambda_1}{\lambda_2 - \lambda_1}\right) \times \ln\left(\frac{\lambda_1}{\lambda_2}\right) \leq \ln(\varepsilon) \quad (19.e)$$

and so we have the optimization problem (19) as follows:

$$\max_{p_{ab},p_j,t} \log_2\left(1+\frac{(1-p_{ab})\gamma_c}{p_{ab}\gamma_c+1}\right)+\ldots$$

$$\left[\begin{array}{l}\log_2\left(1+\dfrac{p_{ab}\gamma_b}{(1-p_{ab})\gamma_b+1}\right)-\ldots\\ \log_2\left(1+\dfrac{p_{ab}\gamma_u}{(1-p_{ab})\gamma_u+p_j\gamma_j+1}\right)\end{array}\right] \quad (19)$$

s.t.

$(19.a),(19.b),(19.c),(19.d)$

$$\left(\frac{\lambda_1}{\lambda_2-\lambda_1}\right)\times\ln\left(\frac{\lambda_1}{\lambda_2}\right)\le\ln(\varepsilon) \quad (19.e)$$

### 4. Solution of optimization problem

As it can be seen, problem (19), and constraints (19.c), (19.d) and (19.e) are not convex, so we cannot use convex optimization solver software such as CVX to solve this problem. In order to make convex the constraint (19.e), an auxiliary variable $t$ will be defined and after applying some mathematical operations on the constraint (19.e), we must solve the following optimization problem:

$$\max_{p_{ab},p_j,t} \log_2\left(1+\frac{(1-p_{ab})\gamma_c}{p_{ab}\gamma_c+1}\right)+\ldots$$

$$\left[\begin{array}{l}\log_2\left(1+\dfrac{p_{ab}\gamma_b}{(1-p_{ab})\gamma_b+1}\right)-\ldots\\ \log_2\left(1+\dfrac{p_{ab}\gamma_u}{(1-p_{ab})\gamma_u+p_j\gamma_j+1}\right)\end{array}\right] \quad (19)$$

s.t.

$(19.a),(19.b),(19.c),(19.d)$

$$\lambda_1\times\ln\left(\frac{\lambda_1}{\lambda_2}\right)-t\times\ln(\varepsilon)\le 0 \quad (19.f)$$

$$\lambda_2-\lambda_1\le t \quad (19.h)$$

We can also use difference of two convex functions (DC) method to make problem (19) and constraints (19.c) and (19.d) convex. We first consider the objective function in optimization problem and so

$$\Xi(p_{ab},p_j)=\Gamma(p_{ab},p_j)-\Omega(p_{ab},p_j) \quad (20)$$

where

$$\begin{cases}\Gamma(p_{ab},p_j)=\log_2(\gamma_c+1)+\log_2(\gamma_b+1)+\cdots\\ \log_2((1-p_{ab})\gamma_u+p_j\gamma_j+1)\\ \\ \Omega(p_{ab},p_j)=\log_2(p_{ab}\gamma_c+1)+\cdots\\ \log_2((1-p_{ab})\gamma_b+1)+\log_2(\gamma_u+p_j\gamma_j+1)\end{cases} \quad (21)$$

using DC method, we can rewrite $\Omega(p_{ab},p_j)$ as follows

$$\begin{cases}\Omega(p_{ab},p_j)\simeq\tilde{\Omega}(p_{ab},p_j)=\Omega(p_{ab}(\mu-1),p_j(\mu-1))\ldots\\ +\nabla^T\Omega(p_{ab}(\mu-1),p_j(\mu-1))\ldots\\ \times[p_{ab}(\mu)-p_{ab}(\mu-1),p_j(\mu)-p_j(\mu-1)]\end{cases} \quad (22)$$

where $\mu$ is the iteration number and $\nabla$ is gradient operator and $\nabla\Omega(p_{ab},p_j)$ is calculated as follows

$$\nabla\Omega(p_{ab},p_j)=$$

$$\left[\begin{array}{c}\dfrac{P_{\max}|h_{ac}|^2 d_{ac}^{-\alpha}}{(\sigma_c^2+p_{ab}P_{\max}|h_{ac}|^2 d_{ac}^{-\alpha})\ln(2)}-\ldots\\ \\ \dfrac{P_{\max}|h_{ab}|^2 d_{ab}^{-\alpha}}{(\sigma_b^2+(1-p_{ab})P_{\max}|h_{ac}|^2 d_{ab}^{-\alpha})\ln(2)},\ldots\\ \\ \dfrac{P_{j\max}|h_{ju}|^2 d_{ju}^{-\alpha}}{(\sigma_u^2+P_{\max}|h_{ac}|^2 d_{ju}^{-\alpha}+p_j P_{j\max}|h_{ju}|^2 d_{ju}^{-\alpha})\ln(2)}\end{array}\right] \quad (23)$$

and objective function of optimization problem can be written as $\Gamma(p_{ab},p_j)-\tilde{\Omega}(p_{ab},p_j)$. As like as objective function of optimization problem, we can use the difference method of two convex functions to approximate the constraints (19.c) and (19.d). Therefor we have constraint (19.c) as follows

$$T(p_{ab}, p_j) - \tilde{\Lambda}(p_{ab}, p_j) \geq 0 \qquad (24)$$

where

$$\begin{cases} T(p_{ab}, p_j) = \log_2(\sigma_b^2 + P_{max}|h_{ab}|^2 d_{ab}^{-\alpha}) + \ldots \\ \log_2(\sigma_u^2 + (1-p_{ab})P_{max}|h_{au}|^2 d_{au}^{-\alpha} + \ldots \\ p_j P_{jmax}|h_{ju}|^2 d_{ju}^{-\alpha}) - R_{Bob}^{min} \end{cases} \qquad (25)$$

$$\begin{cases} \Lambda(p_{ab}, p_j) = \log_2(\sigma_b^2 + P_{max}|h_{ab}|^2 d_{ab}^{-\alpha}) + \cdots \\ \log_2(\sigma_u^2 + P_{max}|h_{au}|^2 d_{au}^{-\alpha} + p_j P_{jmax}|h_{ju}|^2 d_{ju}^{-\alpha}) \end{cases} \qquad (26)$$

and also for the constraint (19.d) we will have:

$$K(p_{ab}, p_j) - \tilde{\Sigma}(p_{ab}, p_j) \geq 0 \qquad (27)$$

where

$$K(p_{ab}, p_j) = \log_2(\sigma_c^2 + P_{max}|h_{ac}|^2 d_{ac}^{-\alpha}) - R_{Carol}^{min} \qquad (28)$$

$$\Sigma(p_{ab}, p_j) = \log_2(\sigma_c^2 + p_{ab} P_{max}|h_{ac}|^2 d_{ac}^{-\alpha}) \qquad (29)$$

It should also be noted that $\tilde{\Lambda}$ and $\tilde{\Sigma}$ are calculated as equation (22). Finally, we can solve the following optimization problem using convex problem-solving toolboxes such as CVX.

$$\max_{p_{ab}, p_j, t} \left( \Gamma(p_{ab}, p_j) - \tilde{\Omega}(p_{ab}, p_j) \right) \qquad (30)$$

s.t.
$(19.a), (19.b), (19.f), (19.h), (24), (27)$

## 5. Incomplete information about Willie's location scenario

As it is mentioned, the eavesdropper (Willie) on the network is passive, so it can be concluded that Alice and the friendly jammer cannot make an accurate estimation on Willie's location. Therefore, in this section, a situation will be considered where accurate information about Willie's location is not available to Alice and the friendly jammer. In other words, it is assumed that although Alice and the friendly jammer have an estimate of their distance from Willie, that is, $\hat{d}_{aw}$ and $\hat{d}_{jw}$, which this estimation is accompanied by an amount of error, which we represent as $e_{d_{aw}} = d_{aw} - \hat{d}_{aw}$ and $e_{d_{jw}} = d_{jw} - \hat{d}_{jw}$, where $e_{d_{aw}}$ and $e_{d_{jw}}$ are estimation error. The distance difference lies in a bounded set i.e. $E_{d_{aw}} = \{e_{d_{aw}} : |e_{d_{aw}}|^2 \leq \tau_{d_{aw}}\}$ and $E_{d_{jw}} = \{e_{d_{jw}} : |e_{d_{jw}}|^2 \leq \tau_{d_{jw}}\}$, where $\tau_{d_{aw}}$ and $\tau_{d_{jw}}$ are known constants. In this case, we should rewrite equations (15) and (16) as follows:

$$p_{FA} = \begin{cases} \left( e^{-\frac{(\theta - \sigma_w^2)}{\lambda_1'}} \right) & \theta - \sigma_w^2 \geq 0 \\ 1 & \theta - \sigma_w^2 < 0 \end{cases} \qquad (31)$$

$$p_{MD} = \begin{cases} \frac{1}{(\lambda_2' - \lambda_1')} \times \left( \lambda_2' \times \left[ 1 - e^{-\frac{(\theta - \sigma_w^2)}{\lambda_2'}} \right] - \cdots \\ \lambda_1' \times \left[ 1 - e^{-\frac{(\theta - \sigma_w^2)}{\lambda_1'}} \right] \right) & \theta - \sigma_w^2 \geq 0 \\ 0 & \theta - \sigma_w^2 < 0 \end{cases} \qquad (32)$$

where $\lambda_1' = p_j P_{jmax}(d_{jw} + \tau_{d_{jw}})^{-\alpha}$ and $\lambda_2' = P_{max}(d_{aw} - \tau_{d_{aw}})^{-\alpha}$.

In this case in order to maximize the average rate of network, following optimization problem is proposed:

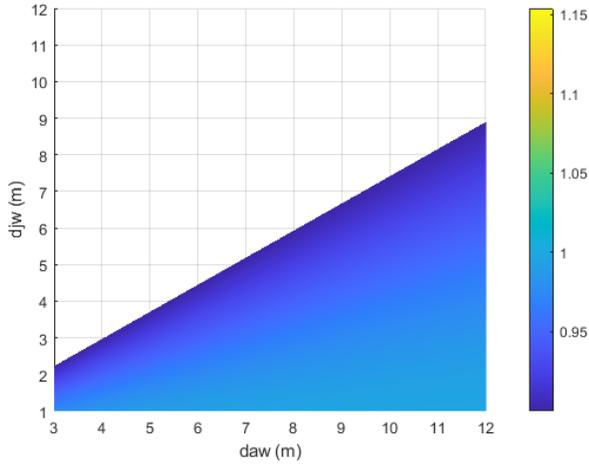

**Figure 2.** Summation of false alarm and missed detection probabilities for different distances of Willie from Alice and friendly jammer.

$$\max_{p_{ab}, p_j, t} \left( \Gamma(p_{ab}, p_j) - \tilde{\Omega}(p_{ab}, p_j) \right) \quad (33)$$

s.t.

$$\left| e_{d_{aw}} \right|^2 \leq \tau_{d_{jw}} \quad (33.a)$$

$$\left| e_{d_{aw}} \right|^2 \leq \tau_{d_{aw}} \quad (33.b)$$

$(19.a), (19.b), (19.f), (19.h), (24), (27).$

In order to solve optimization problem (33), we present following lemma.

**Lemma 1.** Figure 2 has been presented in order to check the satisfaction of the covert communication condition for different distances of Willie from Alice and friendly jammer for this network topology in optimization problem (33). This figure shows the summation of missed detection and false alarm probabilities for Willie's different distances from Alice and the friendly jammer. As can be seen, for this network topology, the covert communication condition is satisfied for different distances from Alice and friendly jammer, and therefore it can be concluded that the imperfect information about Willie's location will not affect the network's performance. At last, to solve optimization problem (33) we can use epigraph and difference of two convex functions methods.

## 6. Numerical results and simulation

In this section, numerical results are presented to evaluate the performance of the proposed network. The simulation parameters in the considered system model are defined in table 1.

**Table 1.** Simulation setting

| | | |
|---|---|---|
| $L_{Alice}$ | Alice's location | (0,0) |
| $L_{Bob}$ | Bob's location | (-10,0) |
| $L_{Carol}$ | Carol's location | (10,0) |
| $L_{UU}$ | Untrusted user's location | (0,10) |
| $L_{Willie}$ | Willie's location | (0,-10) |
| $L_{FJ}$ | Friendly jammer's location | (0,2) |
| $P_{max}$ | Maximum power of Alice | 2 dBW |
| $P_{jmax}$ | Maximum power of friendly jammer | 8 dBW |
| $\sigma_m^2$ | Received noise power at node m | -30 dB |
| $\alpha$ | Path loss exponent | 2 |
| $R_{Bob}^{min}$ | Requested quality of service by Bob [bps/Hz] | 0.2 |
| $R_{Carol}^{min}$ | Requested quality of service by Carol [bps/Hz] | 0.1 |
| $\varepsilon$ | Lower bound of detection error probability at Willie | 0.1 |

Figure 3 shows the effect of Bob's distance from Alice. It is observed that with increasing Bob's distance from Alice, the average rate of network decreases. Since Bob uses both covert communication and information theory security simultaneously, as the Bob's increases from the transmitter, the covert communication condition is not satisfied, and as a result, the transmitter cannot significantly increase the allocated power to Bob, and as a result the network will intelligently devote the rest of its power to Carol in order to increase the average rate. This figure also shows transmitted rates for Bob and Carol separately. As it can be seen, in none of these intervals the rate of none of the users reaches zero, and the minimum quality of service requested by users was also provided.

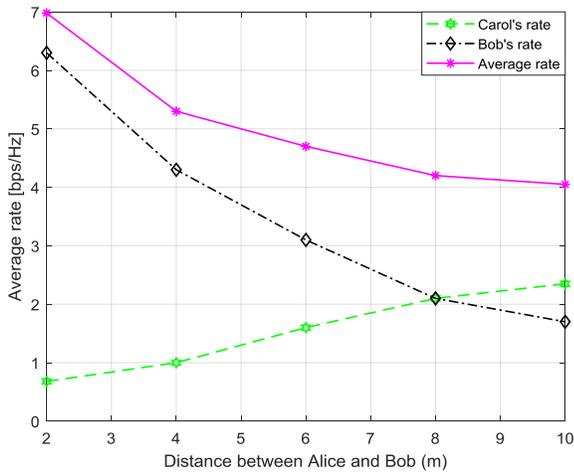

**Figure 3.** The effect of Bob's distance from Alice on average rate.

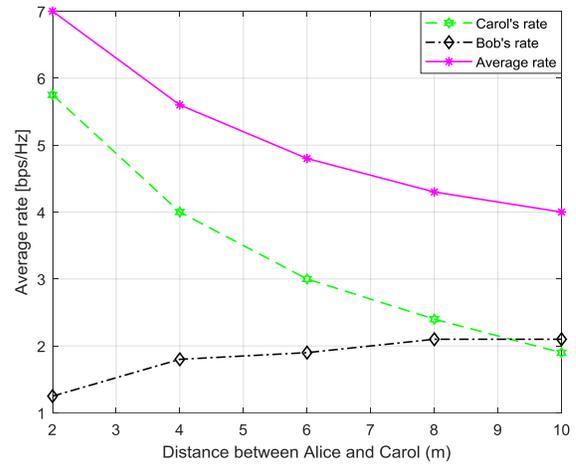

**Figure 4.** The effect of Carol's distance from Alice on average rate.

Figure 4 indicates the effect of Carol's distance from Alice as well as the transmitted rate for each user separately. As can be seen from this figure, the average network's rate decreases with increasing Carol's distance from the transmitter, and at the same time the quality of service requested is provided for both authorized network users. It is also observed that the central controller node of the network intelligently adjusts the power sent to them according to the distance of the users from the transmitter in order to increase the average rate. Similar to the case of increasing Bob's distance from the transmitter, the reason that average rate decreases with increasing Carol's distance from the transmitter is that as Carol moves away from the transmitter, Alice cannot greatly increase the power allocated to Carol

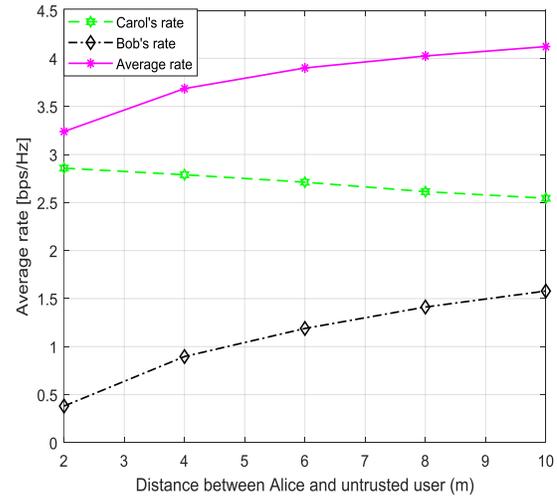

**Figure 5.** The effect of untrusted user's distance from Alice on average rate.

because in such cases, the condition of covert communication is not satisfied.

Figure 5 shows the effect of the untrusted user's distance from the transmitter as well as the sent rate for authorized network users separately. As it can be seen from this figure, the network's average rate has increased dramatically with increasing untrusted user's distance from the transmitter. It is also observed that the central controller node of the network intelligently adjusts the power sent to the authorized users of the network according to the distance of the untrusted user from the transmitter to increase the average rate. As can be seen in this figure, Bob's rate increases significantly with increasing untrusted user's distance.

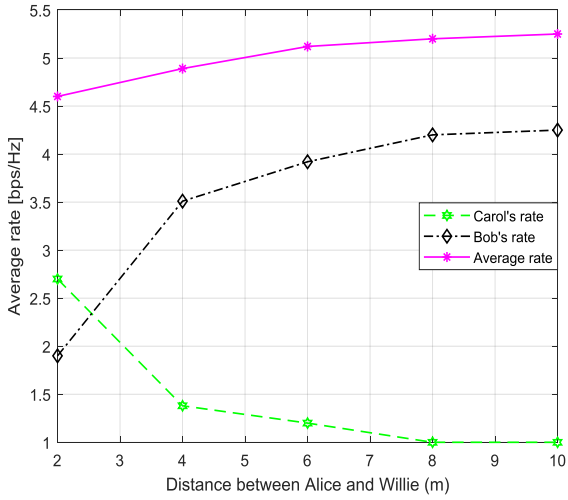

**Figure 6.** The effect of increasing Willie's distance from Alice and friendly jammer.

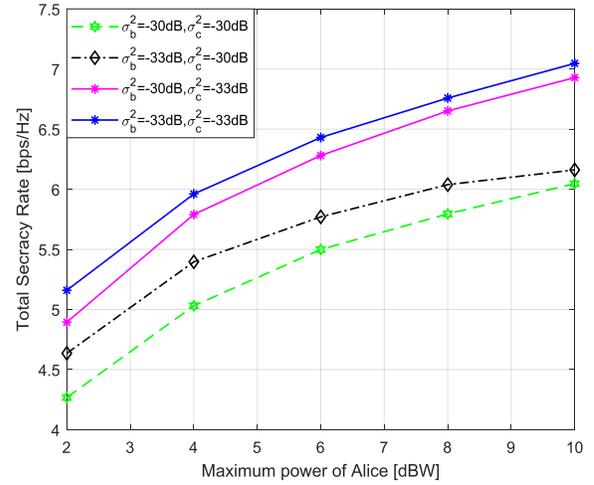

**Figure 7.** The effect of increasing $P_{\max}$ on average rate and $P_{j\max} = 20\,dBW$.

Figure 6 shows the effect of increasing the Willie's distance from Alice and the friendly jammer when the distances of other nodes are assumed to be constant. In this simulation it is assumed the distance of Carol and Bob from Alice is constant and equal to 5 m. As we can see, with increasing Willie's distance from Alice, the average rate increases.

Figure 7 shows the effect of increasing the power of Alice when the friendly jammer's power is assumed to be equal to 20dBW. As can be seen, with the increase of Alice's power from 2dBW to 10dBW, while maintaining the confidentiality of covert and secure transmission, the average rate will increase close to 42%. This diagram also shows the effect of the received noise power on the receiver of authorized users on the average rate. As can be seen, the received noise power at Carol's receiver has a greater effect on the average rate.

## 7. Conclusion

In this article, we examined the security of information theory combined with covert communication according to the different security requirements of users in the network in the presence of friendly jammer. There are two authorized users in the network under investigation, one of which requires secure and covert communication (Bob) and the other requires covert communication (Carol). In this network, it is assumed that the transmitter (Alice) will not have any transmission to any of the two users in a time slot and will send to both authorized users of the network at the same time in another time slot. For the system under investigation, we propose an optimization problem in which our goal is to maximize the average rate according to the requirements of covert communication and information theoretic security, as well as to ensure the quality of service requested by users. Since the optimization problem was not convex, we used the convex approximation method to make the optimization problem convex. The simulation results showed the effect of increasing the distance of authorized network users from the transmitter and also the effect of increasing the transmitter sending power on the average network rate. It was also observed that the network intelligently adjusts the transmission power allocated to each authorized users in order to maximize average rate of network to improve network's performance.

It is noticeable that we examined the scenario of imperfect information about Willie's location in section 5, in this section we assumed Alice and friendly jammer have estimation about Willie's location, but this estimation is not perfect and will have error. In section 5 we showed that this estimation error will not effect on proposed network's performance.

## References

[1] C. S. R. Murthy and B. S. Manoj, *Ad hoc wireless networks: architectures and protocols*. 2004.

[2] Y. S. Shiu, S. Y. Chang, H. C. Wu, S. C. H. Huang, and H. H. Chen, "Physical layer security in wireless networks: A tutorial," *IEEE Wirel. Commun.*, vol. 18, no. 2, pp. 66–74, 2011, doi: 10.1109/MWC.2011.5751298.

[3] N. Yang, S. Yan, J. Yuan, R. Malaney, R. Subramanian, and I. Land, "Artificial noise: Transmission optimization in multi-input single-output wiretap channels," *IEEE Trans. Commun.*, vol. 63, no. 5, 2015, doi: 10.1109/TCOMM.2015.2419634.

[4] M. Bloch and J. Barros, *Physical–layer security: From information theory to security engineering*, vol. 9780521516. 2011.

[5] J. Hu, K. Shahzad, S. Yan, X. Zhou, F. Shu, and J. Li, "Covert Communications with a Full-Duplex Receiver over Wireless Fading Channels," in *IEEE International Conference on Communications*, 2018, vol. 2018-May, doi: 10.1109/ICC.2018.8422941.

[6] A. D. Wyner, "The Wire-Tap Channel," *Bell Syst. Tech. J.*, vol. 54, no. 8, 1975, doi: 10.1002/j.1538-7305.1975.tb02040.x.

[7] T. V. Sobers, B. A. Bash, S. Guha, D. Towsley, and D. Goeckel, "Covert Communication in the Presence of an Uninformed Jammer," in *IEEE Transactions on Wireless Communications*, 2017, vol. 16, no. 9, doi: 10.1109/TWC.2017.2720736.

[8] B. A. Bash, D. Goeckel, D. Towsley, and S. Guha, "Hiding information in noise: Fundamental limits of covert wireless communication," in *IEEE Communications Magazine*, 2015, vol. 53, no. 12, pp. 26–31, doi: 10.1109/MCOM.2015.7355562.

[9] M. R. Bloch, "Covert Communication over Noisy Channels: A Resolvability Perspective," in *IEEE Transactions on Information Theory*, 2016, vol. 62, no. 5, doi: 10.1109/TIT.2016.2530089.

[10] M. Forouzesh, P. Azmi, A. Kuhestani, and P. L. Yeoh, "Joint Information Theoretic Secrecy and Covert Communication in the Presence of an Untrusted User and Warden," *IEEE Internet Things J.*, 2020, doi: 10.1109/JIOT.2020.3038682.

[11] S. Lee, R. J. Baxley, M. A. Weitnauer, and B. Walkenhorst, "Achieving undetectable communication," *IEEE J. Sel. Top. Signal Process.*, vol. 9, no. 7, 2015, doi: 10.1109/JSTSP.2015.2421477.

[12] B. He, S. Yan, X. Zhou, and V. K. N. Lau, "On covert communication with noise uncertainty," *IEEE Commun. Lett.*, vol. 21, no. 4, 2017, doi: 10.1109/LCOMM.2016.2647716.

[13] G. Zheng, I. Krikidis, J. Li, A. P. Petropulu, and B. Ottersten, "Improving physical layer secrecy using full-duplex jamming receivers," *IEEE Trans. Signal Process.*, vol. 61, no. 20, 2013, doi: 10.1109/TSP.2013.2269049.

[14] M. Abedi, N. Mokari, H. Saeedi, and H. Yanikomeroglu, "Secure robust resource allocation using full-duplex receivers," 2015, doi: 10.1109/ICCW.2015.7247229.

[15] M. Duarte, C. Dick, and A. Sabharwal, "Experiment-driven characterization of full-duplex wireless systems," *IEEE Trans. Wirel. Commun.*, vol. 11, no. 12,

2012, doi: 10.1109/TWC.2012.102612.111278.

[16] L. Wang, Y. Cai, Y. Zou, W. Yang, and L. Hanzo, "Joint Relay and Jammer Selection Improves the Physical Layer Security in the Face of CSI Feedback Delays," *IEEE Trans. Veh. Technol.*, vol. 65, no. 8, 2016, doi: 10.1109/TVT.2015.2478029.

[17] D. Bharadia, E. McMilin, and S. Katti, "Full duplex radios," 2013, doi: 10.1145/2486001.2486033.

[18] M. Forouzesh, P. Azmi, and N. Mokari, "Reduce impact of false detection of adversary states on the secure cooperative network," 2017, doi: 10.1109/ISTEL.2016.7881822.

[19] M. Forouzesh, P. Azmi, N. Mokari, and D. Goeckel, "Covert Communication Using Null Space and 3D Beamforming: Uncertainty of Willie's Location Information," *IEEE Trans. Veh. Technol.*, vol. 69, no. 8, pp. 8568–8576, 2020, doi: 10.1109/TVT.2020.2997074.

[20] Y. Wen, Y. Huo, L. Ma, T. Jing, and Q. Gao, "A Scheme for Trustworthy Friendly Jammer Selection in Cooperative Cognitive Radio Networks," *IEEE Trans. Veh. Technol.*, vol. 68, no. 4, 2019, doi: 10.1109/TVT.2019.2895639.

[21] S. Sharifian, F. Lin, and R. Safavi-Naini, "Secret key agreement using a virtual wiretap channel," 2017, doi: 10.1109/INFOCOM.2017.8057119.

[22] K. Cumanan *et al.*, "Physical Layer Security Jamming: Theoretical Limits and Practical Designs in Wireless Networks," *IEEE Access*, vol. 5, 2017, doi: 10.1109/ACCESS.2016.2636239.

[23] K. Cumanan, Z. Ding, M. Xu, and H. V. Poor, "Secure multicast communications with private jammers," in *IEEE Workshop on Signal Processing Advances in Wireless Communications, SPAWC*, 2016, vol. 2016-August, doi: 10.1109/SPAWC.2016.7536824.

[24] L. Wang, M. Elkashlan, J. Huang, N. H. Tran, and T. Q. Duong, "Secure transmission with optimal power allocation in untrusted relay networks," *IEEE Wirel. Commun. Lett.*, vol. 3, no. 3, 2014, doi: 10.1109/WCL.2014.031114.140018.